\documentclass[12pt]{article}
\pdfoutput=1
\usepackage[colorlinks,linkcolor=Blue,citecolor=Blue,bookmarks,bookmarksnumbered]{hyperref}
\usepackage[scaled=0.85]{helvet}
\usepackage{amsmath,amssymb,accents,mathrsfs,XoohmE}
\usepackage{graphicx,color}
\usepackage{booktabs}
\usepackage{multirow}
\usepackage{placeins}
\usepackage{amsmath}

\usepackage{XoohmE}

\definecolor{Green}  {rgb}{0.10,0.70,0.10} 
\definecolor{Orange} {rgb}{1.00,0.50,0.15} 
\definecolor{Red}    {rgb}{0.90,0.00,0.12} 
\definecolor{Purple} {rgb}{0.50,0.25,0.55} 
\definecolor{Turque} {rgb}{0.00,0.65,0.85} 
\definecolor{Blue}   {rgb}{0.00,0.00,1.00} 
\definecolor{Magenta}{rgb}{1.00,0.00,1.00} 
\definecolor{Gold}   {rgb}{1.00,0.75,0.25} 
\definecolor{Seaweed}{rgb}{0.01,0.24,0.09} 
\definecolor{Brown}  {rgb}{0.43,0.26,0.32} 
\definecolor{grey1}  {rgb}{0.20,0.20,0.20} 
\definecolor{grey2}  {rgb}{0.40,0.40,0.40} 
\definecolor{grey3}  {rgb}{0.60,0.60,0.60} 
\definecolor{grey4}  {rgb}{0.80,0.80,0.80} 
\definecolor{grey5}  {rgb}{0.90,0.90,0.90} 
\def\C#1#2{{\ifcase#1\or
             \color{Green}\or \color{Orange}\or \color{Red}\or
              \color{Purple}\or \color{Turque}\or \color{Blue}\or
               \color{Magenta}\or \color{Gold}\or \color{Seaweed}\or
                \color{Brown}\or\color{grey1}\or\color{grey2}\or
                 \color{grey3}\else\color{grey4}\fi#2}}

\definecolor{Slate} {rgb}{0.00,0.45,0.55}

\def\rD{{\rm D}}

\def\fracm#1#2{\hbox{\large{${\frac{{#1}}{{#2}}}$}}}

\def\be{\begin{equation}}
\def\ee{\end{equation}}
\newcommand{\bea}{\begin{eqnarray}}
\newcommand{\eea}{\end{eqnarray}}
\newcommand{\ena}{\end{eqnarray}}

\def\pp{{\mathchoice
          {
              \kern 1pt%
              \raise 1pt
              \vbox{\hrule width5pt height0.4pt depth0pt
                    \kern -2pt
                    \hbox{\kern 2.3pt
                          \vrule width0.4pt height6pt depth0pt
                          }
                    \kern -2pt
                    \hrule width5pt height0.4pt depth0pt}%
                    \kern 1pt
           }
            {
              \kern 1pt%
              \raise 1pt
              \vbox{\hrule width4.3pt height0.4pt depth0pt
                    \kern -1.8pt
                    \hbox{\kern 1.95pt
                          \vrule width0.4pt height5.4pt depth0pt
                          }
                    \kern -1.8pt
                    \hrule width4.3pt height0.4pt depth0pt}%
                    \kern 1pt
            }
            {
              \kern 0.5pt%
              \raise 1pt
              \vbox{\hrule width4.0pt height0.3pt depth0pt
                    \kern -1.9pt  
                    \hbox{\kern 1.85pt
                          \vrule width0.3pt height5.7pt depth0pt
                          }
                    \kern -1.9pt
                    \hrule width4.0pt height0.3pt depth0pt}%
                    \kern 0.5pt
            }
            {
              \kern 0.5pt%
              \raise 1pt
              \vbox{\hrule width3.6pt height0.3pt depth0pt
                    \kern -1.5pt
                    \hbox{\kern 1.65pt
                          \vrule width0.3pt height4.5pt depth0pt
                          }
                    \kern -1.5pt
                    \hrule width3.6pt height0.3pt depth0pt}%
                    \kern 0.5pt
            }
        }}

\def\mm{{\mathchoice
                       {
                             \kern 1pt
               \raise 1pt    \vbox{\hrule width5pt height0.4pt depth0pt
                                  \kern 2pt
                                  \hrule width5pt height0.4pt depth0pt}
                             \kern 1pt}
                       {
                            \kern 1pt
               \raise 1pt \vbox{\hrule width4.3pt height0.4pt depth0pt
                                  \kern 1.8pt
                                  \hrule width4.3pt height0.4pt depth0pt}
                             \kern 1pt}
                       {
                            \kern 0.5pt
               \raise 1pt
                            \vbox{\hrule width4.0pt height0.3pt depth0pt
                                  \kern 1.9pt
                                  \hrule width4.0pt height0.3pt depth0pt}
                            \kern 1pt}
                       {
                           \kern 0.5pt
             \raise 1pt  \vbox{\hrule width3.6pt height0.3pt depth0pt
                                  \kern 1.5pt
                                  \hrule width3.6pt height0.3pt depth0pt}
                           \kern 0.5pt}
                       }}

\def\ad{{\kern0.5pt
                   \alpha \kern-5.05pt \raise5.8pt\hbox{$\textstyle.$}\kern
0.5pt}}

\def\bd{{\kern0.5pt
                   \beta \kern-5.05pt \raise5.8pt\hbox{$\textstyle.$}\kern
0.5pt}}

\def\qd{{\kern0.5pt
                   q \kern-5.05pt \raise5.8pt\hbox{$\textstyle.$}\kern
0.5pt}}
\def\Dot#1{{\kern0.5pt
     {#1} \kern-5.05pt \raise5.8pt\hbox{$\textstyle.$}\kern
0.5pt}}

\catcode`@=11
\def\un#1{\relax\ifmmode\@@underline#1\else
        $\@@underline{\hbox{#1}}$\relax\fi}
\catcode`@=12

\def\a{\alpha}
\def\b{\beta}
\def\c{\chi}
\def\d{\delta}
\def\e{\epsilon}

\def\g{\gamma}

\def\i{\iota}
\def\j{\psi}
\def\k{\kappa}
\def\l{\lambda}
\def\m{\mu}
\def\n{\nu}
\def\o{\omega}

\def\r{\rho}
\def\s{\sigma}
\def\t{\tau}

\def\L{\Lambda}
\def\O{\Omega}
\def\P{\Pi}

\def\S{\Sigma}

\def\ca{{\cal A}}

\def\cf{{\cal F}}

\def\dslash{\not{\hbox{\kern-2pt $\partial$}}}
\def\Dslash{\not{\hbox{\kern-4pt $D$}}}
\def\pslash{\not{\hbox{\kern-2.3pt $p$}}}
 \newtoks\slashfraction
 \slashfraction={.13}
 \def\slash#1{\setbox0\hbox{$ #1 $}
 \setbox0\hbox to \the\slashfraction\wd0{\hss \box0}/\box0 }
 
\def\kcr{{\hbox{\ro \char'170}}}                
\def\ktl{{\hbox{\ro \char'170}}}        
\def\ktr{{\hbox{\ro \char'170}}}        
\def\kbl{{\hbox{\ro \char'170}}}        
\def\kbr{{\hbox{\ro \char'170}}}        

\def\plpl{\raise-2pt\hbox{$\raise3pt\hbox{$_+$}\hskip-6.67pt\raise0.0pt
\hbox{$^+$}\hskip 0.01pt$}}
\def\mimi{\raise-2pt\hbox{$\raise3pt\hbox{$_-$}\hskip-6.67pt\raise0.0pt
\hbox{$^-$}\hskip 0.01pt$}} 

\def\bo{{\raise.15ex\hbox{\large$\Box$}}}               
\def\pa{\partial}                                       
\def\TH{{\raise.2ex\hbox{$\displaystyle \bigodot$}\mskip-4.7mu \llap H \;}}
\def\face{{\raise.2ex\hbox{$\displaystyle \bigodot$}\mskip-2.2mu \llap {$\ddot
        \smile$}}}                                      

\def\dt#1{\on{\hbox{\bf .}}{#1}}                
\def\Dot#1{\dt{#1}}

\def\Tilde#1{\widetilde{#1}}                    
\def\Hat#1{\widehat{#1}}                        
\def\leftrightarrowfill{$\mathsurround=0pt \mathord\leftarrow \mkern-6mu
        \cleaders\hbox{$\mkern-2mu \mathord- \mkern-2mu$}\hfill
        \mkern-6mu \mathord\rightarrow$}
\def\dvec#1{\vbox{\ialign{##\crcr
        \leftrightarrowfill\crcr\noalign{\kern-1pt\nointerlineskip}
        $\hfil\displaystyle{#1}\hfil$\crcr}}}           
\def\dt#1{{\buildrel {\hbox{\LARGE .}} \over {#1}}}     

\def\fracm#1#2{\hbox{\large{${\frac{{#1}}{{#2}}}$}}}
\def\sfrac#1#2{{\vphantom1\smash{\lower.5ex\hbox{\small$#1$}}\over
        \vphantom1\smash{\raise.4ex\hbox{\small$#2$}}}} 
\def\bfrac#1#2{{\vphantom1\smash{\lower.5ex\hbox{$#1$}}\over
        \vphantom1\smash{\raise.3ex\hbox{$#2$}}}}       
\def\afrac#1#2{{\vphantom1\smash{\lower.5ex\hbox{$#1$}}\over#2}}    

\def\pa{\partial}      
\let\bm\relax
\newcommand{\bm}[1]{{\boldsymbol{#1}}}

\def\ad{{\dot{\alpha}}}
\def\bd{{\dot{\beta}}}

 \font\rOpe=cmsy10                        
 \def\ktl{{\hbox{\rOpe\char'170}}}        
 \def\kbl{{\hbox{\rOpe\char'170}}}        
 \def\kcr{{\reflectbox{\rOpe\char'170}}}        
 \def\ktr{{\reflectbox{\rOpe\char'170}}}        
 \def\kbr{{\reflectbox{\rOpe\char'170}}}        
 \def\Border{\vbox{\hsize0pt
        \setlength{\unitlength}{1mm}
        \newcount\xco
        \newcount\yco
        \xco=-21
        \yco=12
        \begin{picture}(0,0)(-7.5,0)
        \put(\xco,\yco){$\ktl$}
        \advance\yco by-1
        {\loop
        \put(\xco,\yco){$\kcr$}
        \advance\yco by-2
        \ifnum\yco>-240
        \repeat
        \put(\xco,\yco){$\kbl$}}
        \xco=170
        \yco=12
        \put(\xco,\yco){$\ktr$}
        \advance\yco by-1
        {\loop
        \put(\xco,\yco){$\kcr$}
        \advance\yco by-2
        \ifnum\yco>-240
        \repeat
        \put(\xco,\yco){$\kbr$}}
        \put(-19.5,13){\scalebox{.6065}{%
         University of Maryland Center for String and Particle  Theory \&\ Physics Department%
        |University of Maryland Center for String and Particle  Theory \&\ Physics Department}}
        \put(-19.5,-241.5){\scalebox{.5835}{%
         ****University of Maryland * Center for String and
         Particle  Theory* Physics Department****University of Maryland *Center
        for String and Particle  Theory* Physics Department}}
        \end{picture}
        \par\vskip-8mm}}
\definecolor{UMred}{rgb}{.9,.05,.2}
\definecolor{HUblue}{rgb}{.0,.3,.7}

\definecolor{Red}    {rgb}{0.90,0.00,0.12} 
\definecolor{Blue}   {rgb}{0.00,0.00,1.00} 
\definecolor{Green}  {rgb}{0.10,0.70,0.10} 
\definecolor{Turque} {rgb}{0.00,0.65,0.85} 
\definecolor{Orange} {rgb}{1.00,0.50,0.15} 
\definecolor{Magenta}{rgb}{1.00,0.00,1.00} 
\definecolor{Gold}   {rgb}{1.00,0.75,0.25} 
\definecolor{Seaweed}{rgb}{0.01,0.24,0.09} 
\definecolor{Purple} {rgb}{0.50,0.25,0.55} 
\definecolor{Brown}  {rgb}{0.43,0.26,0.32} 
\definecolor{grey1}  {rgb}{0.20,0.20,0.20} 
\definecolor{grey2}  {rgb}{0.40,0.40,0.40} 
\definecolor{grey3}  {rgb}{0.60,0.60,0.60} 
\definecolor{grey4}  {rgb}{0.80,0.80,0.80} 
\definecolor{grey5}  {rgb}{0.90,0.90,0.90} 
\def\C#1#2{{\ifcase#1\or
             \color{Red}\or \color{Green}\or \color{Blue}\or\
              \color{Turque}\or \color{Orange}\or \color{Magenta}\or 
               \color{Gold}\or \color{Seaweed}\or \color{Purple}\or
                \color{Brown}\or\color{grey1}\or\color{grey2}\or
                 \color{grey3}\else\color{grey4}\fi#2}}

\definecolor{Slate} {rgb}{0.00,0.45,0.55}

\newdimen\parshift\parshift=\parindent
\catcode`@=11
 \long\def\@footnotetext#1{\insert\footins{\reset@font\footnotesize
           \interlinepenalty\interfootnotelinepenalty\splittopskip%
            \footnotesep\splitmaxdepth\dp\strutbox\floatingpenalty\@MM%
             \hsize\columnwidth\addtolength{\hsize}{-2\parindent}
              \@parboxrestore\protected@edef\@currentlabel%
              {\csname p@footnote\endcsname\@thefnmark}%
                \color@begingroup%
                 \@makefntext{\rule\z@\footnotesep\ignorespaces#1%
                  \@finalstrut\strutbox}%
                \color@endgroup}}
 \long\def\@makefntext#1{\hglue\parshift%
           \vbox{\noindent\baselineskip=11pt plus.5pt minus.5pt\hb@xt@0em{\hss\@makefnmark\kern1pt}#1}}
\catcode`@=12

\newskip\humongous \humongous=0pt plus 1000pt minus 1000pt
\def\caja{\mathsurround=0pt}
\def\eqalign#1{\,\vcenter{\openup2\jot \caja
        \ialign{\strut \hfil$\displaystyle{##}$&$
        \displaystyle{{}##}$\hfil\crcr#1\crcr}}\,}
\newif\ifdtup

\makeatletter
\def\section{\@startsection{section}{1}{\z@}
        {3ex plus-1ex minus-.2ex}{1pt plus1pt}{\large\sf\bfseries\boldmath}}
\def\subsection{\@startsection{subsection}{2}{\z@}
         {1.5ex plus-1ex minus-.2ex}{0.01pt plus1pt}{\sf\slshape}}
\def\subsubsection{\@startsection{subsubsection}{3}{\z@}
          {1.5ex plus-1ex minus-.2ex}{0.01pt plus0.2pt}{\sf\boldmath}}
\def\paragraph{\@startsection{paragraph}{4}{\z@}
           {.75ex \@plus.5ex \@minus.2ex}{-2mm}{\sf\bfseries\boldmath}}
\makeatother

 \allowdisplaybreaks
 \seceq

\begin{document}

\thispagestyle{empty}
\noindent{\small
\hfill{HET-1789}  \\ 
{}
}
\begin{center}
{\large \bf
On the Ubiquity Of Electromagnetic-Duality Rotations in \vskip0.1in
4D, $\mathcal{N}=1$ Holoraumy Tensors for On-Shell 4D Supermultiplets
}   \\   [4mm]
{\large {
S.\ James Gates, Jr.,\footnote{sylvester$_-$gates@brown.edu}${}^{,a, b}$
Daniel Lay,\footnote{dlay@terpmail.umd.edu}${}^{, c}$ 
S.-N. Hazel Mak,\footnote{sze$_-$ning$_-$mak@brown.edu}${}^{,b}$
Brock Peters,\footnote{bpeters@terpmail.umd.edu}${}^{, c}$ 
Aravind Ramakrishnan,\footnote{akr947@umd.edu}${}^{, c}$ 
Kory Stiffler,\footnote{kory$_-$stiffler@brown.edu}${}^{,a, b}$
Zachary Wimpee,\footnote{zwimpee@angelo.edu}${}^{, d}$
Xiao Xiao,\footnote{1155091983@link.cuhk.edu.hk}${}^{, e}$
Yifan Yuan,\footnote{yuano77@terpmail.umd.edu}${}^{, c}$ 
Jinjie Zhang,\footnote{jzhang54@terpmail.umd.edu}${}^{, c}$ 
Peter V. Zhou\footnote{pzhou1@umd.edu}${}^{, c}$
}}
\\*[6mm]
\emph{
\centering
$^{a}$Brown Theoretical Physics Center,
\\[1pt]
$^{b}$Department of Physics, Brown University,
\\[1pt]
Box 1843, 182 Hope Street, Barus \& Holley 545,
Providence, RI 02912, USA 
\\[10pt]
$^{c}${Center for String and Particle Theory-Dept.\ of Physics, University of Maryland, \\ 4150 Campus Dr., College Park, MD 20472,  USA}
\\[10pt]
$^{d}$Angelo State University, 2601 W. Avenue N, San Angelo, Texas 76909, USA
\\[4pt]
and
\\[4pt]
$^{e}$The Chinese University of Hong Kong, Shatin, New Territories, Hong Kong, China
}
 \\*[5mm]
{ ABSTRACT}\\[5mm]
\parbox{142mm}{\parindent=2pc\indent\baselineskip=14pt plus1pt
Holoraumy is a tool being developed for dimensional enhancement 
(supersymmetry holography) where the goal is to build higher 
dimensional supersymmetric multiplets from lower dimensional 
supersymmetric multiplets.  In this paper, for the first time we investigate 
holoraumy for on-shell supersymmetry. Specifically, the holoraumy 
tensors for a number of familiar 4D, $\mathcal{N}=1$ multiplets are calculated.  It is shown in all of these
cases of on-shell theories, the holoraumy is of the form of an electromagnetic
duality charge multiplying a composite transformation involving an 
electromagnetic duality rotation through an angle of $\pi/2$ times
a space time translation. The details of our calculations can be found at the HEPTHools \href{https://hepthools.github.io/Data/}{Data Repository}.
}
 \end{center}
\vfill
\vfill
\clearpage
%

\newpage
\section{Introduction}

In the works of \cite{4dHolor1,4dHolor2,4dHolor3}, the concept of ``holoraumy'' in four 
dimensions\footnote{The interested reader can find in the bibliography of these works
the citations for introductions of this concept into SUSY QM models and the physics
literature.} was introduced.  This occurred as an outgrowth of developments surrounding 
minimal adinkras related to the Coxeter Group BC${}_4$\cite{1dHolor1,1dHolor2,permutadnk}. In 
this latter context it was noted the transport of any fermionic adinkra node that occurs in 
a valise adinkra under the action of two successive applications of supercharges in 
comparison to the application of the supercharges in the opposite order acting on the 
fermionic nodes leads to either a rotation times a one-dimensional translation or an 
extended R-symmetry transformation times a one-dimensional translation.  

The appearance \cite{4dHolor1,4dHolor2,4dHolor3} of this rotation was noted as a 
rough analogy to the behavior of parallel transport of a vector around a closed path 
in the differential geometry of a higher dimensional Riemannian manifold.  In the 
case for such a D-dimensional manifold such rotations appear and when these 
are considered over everytwo-symplex, one is led to a definition of the Riemann 
curvature tensor and an intrinsic characterization of the local curvature of the manifold.

Since ultimately, the whole goal of the adinkra program is to achieve a more rigorous
understanding of SUSY representations in all dimensions, our research program
moves alternately back-\&-forth between one dimensional adinkra based investigations
to the domains on supersymmmetrical theories in higher dimensions.  The current
work will probe a question not under previous investigation.  Namely the works in
\cite{4dHolor1,4dHolor2}, concentrated on ``off-shell'' theories, but no investigations 
have been carried out into the structure of holoraumy in the absence of a complete
set of auxiliary fields in four dimensional SUSY theories, i.\ e.\ ``on-shell'' theories.  
This is the question we pursue in the current work. 

In the second chapter of this work, we simply review the spectrum and supersymmetry
variations of the: \newline
\indent
(a.) chiral and/or complex linear supermultiplets\footnote{The on-shell spectrum as 
well as the SUSY variations of the chiral and/or complex linear supermultiplets are 
identical.  Thus, it is not possible to distinguish one from the other in the limit where all 
auxiliary fields have been removed.}, \newline
\indent
(b.) vector supermultiplet,
\newline
\indent
(c.)  axial- vector supermultiplet,
\newline
\indent
(d.) matter gravitino supermultiplet, and
\newline
\indent
(e.) supergravity supermultiplet.

The third chapter begins with a general discussion of the methodology for obtaining
the form of the SUSY closure algebra for any of the supermultiplets discussed in
the previous chapter.  Using this methodology on each of the supermultiplets in 
the second chapter allows for a uniform derivation of non-closure functions in a 
systematic manner.  These results have been computed many times and in many 
places previously.  For our purposes, it is simply convenient to provide them (along 
with derivations) for the reader.

The fourth chapter reviews holoraumy for 4D off-shell multiplets. The fifth chapter
is the beginning of presentations of results we believe have not been discussed 
previously in the literature for on-shell four dimensional theories.  We note uniformly 
on each of the supermultiplets, the on-shell theories all involve several common 
structures.  One of these includes a uniform appearance of an ``electromagnetic 
duality rotation" implemented on both the bosons and fermion components when 
calculating the on-shell holoraumy.  When one looks at supermultiplets that contain 
gauge bosons the electromagnetic duality rotation is precisely the usual one effecting 
``rotations'' of electric and magnetic fields.  The holoraumy tensors all utilize an angle 
of $\pi$/2 and depend on a electromagnetic duality charge.  This charge is strictly 
determined by the difference of two of the integer parameters that characterize the 
off-shell holoraumy.  We find in on-shell theories that upon enforcing the equations 
of motion the information contained in the sum of these two parameters, along with 
another two integer parameters is totally ``lost.''

The sixth chapter includes our conclusions.  We include one appendix that gives 
explicit expressions for the various algebraic structure introduced and used in 
our calculations.

\newpage
\section{4D, \texorpdfstring{$\mathcal{N}=1$}{N=1} On-shell Supermultiplets}

The structure of on-shell supermultiplets (i.e. ones in which the supersymmtery algebra
closes {\em {only}} with the use of supplementary kinematic conditions) is widely known.
With the notable exception of the 4D, $\cal N$ = 1 double tensor supermultiplet \cite{Frdmn},
it is known how to introduce ``auxiliary fields'' for these theories in such a way so that
these supplementary conditions can be avoided and one obtains ``off-shell'' representations.
Since the ``auxiliary fields'' for the 4D, $\cal N$ = 1 double tensor supermultiplet
are not known, it is the exception rather than the rule for such theories.

It is also the case when the auxiliary fields are unknown, then it follows the theory in 
question {\em {cannot}} be expressed in terms of a set of {\em {unconstrained}} 
superfields with the attributes of being free from any a priori constraints.  In 
the context of the quantization for non-supersymmetrical electromagnetism, this 
is precisely the reason that one shifts from the formulation being written in terms 
of the electric and magnetic fields to a formulation expressed in terms of the 
four-potential as the latter is free of such a priori constraints.  This emphasizes in 
the context of electromagnetism the formulation of the theory in terms of the 
``E-fields'' and ``B-fields'' constitutes the ``on-shell'' formulation while the formulation 
in terms of the ``A-field'' describes the ``off-shell'' formulation.  

It is the role of the dynamical restrictions on the fields generically that is so strikingly different
for these most interesting theories that suggests there remain unanswered and largely
unexplored questions about the structure of supersymmetrical field theories and 
beyond.  

In this section we will review in one place the results that are known for the ``on-shell''
closure of the supersymmetry algebra for the cases of supermultiplets.  The equations
in (2.1) - (2.5) concisely present the case of the 4D, $\cal N$ = 1 supermultiplets in
their on-shell formulation for spins less than two.

\subsection{Chiral Supermultiplet and Complex Linear Supermultiplet On-Shell 
\texorpdfstring{$(A,B,\psi_{c})$}{A,B,jc}}
\begin{align}
\begin{split}
 \text{D}_{a} A ~=&~ \psi_a ~~,~~  \text{D}_{a} B ~=~
i (\g^{5})_{a}{}^{b} \psi_{b}  ~~,~~ 
\text{D}_{a} \psi_{b} ~=~ i (\g^{\m})_{ab} \pa_{\m} A - (\g^{5}
\g^{\m})_{ab} \pa_{\m} B ~~~.
\end{split}
\label{minSM1}
\end{align}

\subsection{Vector Supermultiplet \texorpdfstring{$(A_{\m},\l_{c})$}{Am,lc}}
\begin{align}
\begin{split}
 \text{D}_{a} A_{\m} ~=&~ (\g_{\m})_{a}{}^{b} \l_{b} ~~,~~
 \text{D}_{a} \l_{b} ~=~ -i \tfrac{1}{2} ([\g^{\m},\g^{\n}])_{ab} 
 \pa_{\m} A_{\n} ~~.
\end{split}
\label{minSM2}
\end{align}

\subsection{Axial-Vector Supermultiplet \texorpdfstring{$(U_{\m},\tilde{\l}_{c})$}{Um,lc}}
\begin{align}
\begin{split}
 \text{D}_{a} U_{\m} ~=&~ i(\g^{5}\g_{\m})_{a}{}^{b} \tilde{\l}_{b} 
 ~~,~~
 \text{D}_{a} \tilde{\l}_{b} ~=~  \tfrac{1}{2} (\g^{5}[\g^{\m},
 \g^{\n}])_{ab} \pa_{\m} U_{\n} ~~.
\end{split}
\label{minSM3}
\end{align}

\subsection{Matter-Gravitino Multiplet \texorpdfstring{$(B_{\m},\psi_{\m c})$}{Bm,pmc}}
\begin{align}
\begin{split}
\text{D}_{a} B_{\m} ~=&~ \psi_{\m a}  ~~,~~ \text{D}_{a} \psi_{\m b} ~=~  - i \tfrac{1}{2} 
 (\g_{\m}[\g^{\a},\g^{\b}])_{ab} \pa_{\a} B_{\b}
\end{split}
\label{minSM4}
\end{align}

\subsection{Supergravity Multiplet \texorpdfstring{$(h_{\m\n},\psi_{\m c})$}{hmn,pmc}}
\begin{align}
\begin{split}
 \text{D}_{a} h_{\m\n} ~=&~ \tfrac{1}{2} (\g_{(\m})_{a}{}^{b} \psi_{\n) b}  ~~,~~
\text{D}_{a} \psi_{\m b} ~=~  - i \tfrac{1}{2} ([\g^{\a},\g^{\b}])_{ab} 
\pa_{\a} h_{\b\m} ~~.
\end{split}
\label{minSM5}
\end{align}
where we note $h_{\m\n}=h_{\n\m}$ and we have made used the notational
convention $A_{(\m} B_{\n)} \equiv A_{\m} B_{\n} + A_{\n} B_{\m}$.

We have not included the discussion of the 4D, $\cal N$ = 1 tensor supermultiplet as
the fields and their SUSY transformation laws in both the on-shell and off-shell cases 
are exactly the same.

\newpage
\section{4D, \texorpdfstring{$\cal N$}{N} = 1 Supersymmetry Algebra}

In this section we place the review of all the SUSY commutator algebras in the same
place on a common basis for all the on-shell supermultiplets of the previous chapter.

To calculate the fermionic SUSY algebra of the $\text{D}_{a}$ operators, we 
expand the anti-commutator algebra with itself in the basis of symmetrical matrices
$ (\g^{\rho})_{ab}$ and $([\g^{\k},\g^{\l}])_{ab}$
\begin{equation}
 \{\text{D}_{a},\text{D}_{b}\} = (\g^{\rho})_{ab} \, \mathcal{X}_{\rho} ~+~ 
 ([\g^{\k},\g^{\l}])_{ab} \, \mathcal{Y}_{\k\l}
\end{equation}
in terms of quantities $\mathcal{X}_{\rho} $ and $\mathcal{Y}_{\k\l}$.  
Since we have the identities
\begin{equation}
\begin{aligned}
 (\g^{\s})^{ab}(\g^{\rho})_{ab} ~=&~ - 4 \eta^{\s\rho} ~~~,  \\
 (\g^{\s})^{ab}([\g^{\k},\g^{\l}])_{ab} ~=&~ 0 ~~~,  \\
([\g^{\rho},\g^{\s}])^{ab}([\g^{\k},\g^{\l}])_{ab} ~=&~ 
16 ( \eta^{\rho\k} \eta^{\s\l} - \eta^{\rho\l} \eta^{\s\k} )
~~,
\end{aligned}
\end{equation}
it follows we may write
\begin{align}
\mathcal{X}_{\rho} =& - \frac{1}{4} (\g_{\rho})^{ab} \{\text{D}_{a},\text{D}_{b}\} 
= - \frac{1}{2} (\g_{\rho})^{ab} \text{D}_{a}\text{D}_{b} ~~,  \\
\mathcal{Y}_{\k\l} =& \frac{1}{32} ([\g_{\k},\g_{\l
}])^{ab} \{\text{D}_{a},\text{D}_{b}\} = \frac{1}{16} ([\g_{\k},\g_{\l
}])^{ab} \text{D}_{a}\text{D}_{b} ~~.
\end{align}
The two operators on the far right hand sides of (3.3) and (3.4) can be evaluated
on and of the fields in (\ref{minSM1}) - (\ref{minSM5}) by iteration of the $\text{D}_{a}$ operators.

\subsection{Chiral Supermultiplet and Complex Linear Supermultiplet On-Shell \texorpdfstring{$(A,B,\psi_{c})$}{A,B,pc}}

Given the transformation laws for the fields in this supermultiplet, we obtain the 4D algebra
given below.
\begin{align}
\{ \text{D}_a , \text{D}_b \} A ~=&~ i 2 (\g^\m)_{ab} \pa_\m A   ~~,~~
\{ \text{D}_a , \text{D}_b \} B ~=~ i 2 (\g^\m)_{ab} \pa_\m B ~~,  \\
\label{e:CMpsiAlgebra}
\{ \text{D}_a , \text{D}_b \} \psi_c ~=&~ i (\g^\m)_{ab}\pa_\m\psi_c 
~-~ i \tfrac{1}{2} (\g^\rho)_{ab}([\g_\rho,\g^\m])_c{}^d\pa_\m\psi_d  \cr
~=&~ i 2  (\g^\m)_{ab}\pa_\m\psi_c - i  (\g^\rho)_{ab}(\g_\rho
\g^\m)_c{}^d\pa_\m\psi_d \cr
~=&~ i 2  (\g^\m)_{ab}\pa_\m\psi_c ~-~ Z^{(CS)}_{abc}~~~.
\end{align}
To pass from the first line of Eq.~(\ref{e:CMpsiAlgebra}) we have made use of the
following identity:
\begin{align}\label{e:Fierz1}
(\g^\rho)_{ab} ([\g_\rho, \g^\m])_{cd} ~=~ 2 (\g^\rho)_{ab} (\g_\rho \g^\m)_{cd}- 2 (\g^\m)_{ab} C_{cd} ~~.
\end{align}
The final term in Eq.~(\ref{e:CMpsiAlgebra}) 
\begin{align}
Z^{(CS)}_{abc}&= i  (\g^\rho)_{ab}(\g_\rho\g^\m)_c{}^d \pa_\m \psi_d
\end{align}
has a couple of descriptors in the literature.  In some places, it is called ``an off-shell central charge''
and in others is referred to as the ``non-closure function'' in the SUSY algebra.  It can be seen the off-shell
central charge $Z^{(CS)}_{abc}$ vanishes upon enforcing the equation of motion for the fermion 
$i(\g^\m)_c{}^d \pa_\m \psi_d  = 0$ as it must.

It is of note, however, that in more complicated supersymmetrical theories, there are known
cases \cite{onSZ} where central charges occur in the SUSY algebra, but that these do not possess 
the property of vanishing when the equations of motion are enforced.  To distinguish such
cases, these are known as ``on-shell central charges.''

\subsection{Vector Supermultiplet \texorpdfstring{$(A_{\m},\l_{c})$}{Am,lc}}

Turning next to the vector supermultiplet, we obtain
\begin{align}
\{ \text{D}_a , \text{D}_b \} A_\m ~=&~ i 2 (\g^\n)_{ab} \pa_{[\n} A_{\m]} \cr
~=&~ i 2 (\g^\n)_{ab} \pa_{\n} A_{\m} - \pa_\m \l^{(VS)}_{ab}
~~~,~~~\l^{(VS)}_{ab} = i 2 (\g^\n)_{ab}A_{\n}  ~~, 
\label{susyMax}
\end{align}
for the vector potential $A{}_{\m}$-field.  We also see the characteristic appearance of
a term that (when contracted with two SUSY parameters) describes a gauge transformation.
The appearance of such terms is ubiquitous in supermultiplets that contain gauge fields.
Evaluation of the anti-commutator on the spinorial field in the supermultiplet leads via
a direct calculation and use of a Fierz identity to the result below
\begin{align}
\begin{split}
\{ \text{D}_a , \text{D}_b \} \l_c ~=&~ i \Big[ \tfrac{3}{2} (\g^\m)_{ab} 
\delta_c{}^d - \tfrac{1}{4} (\g^\n)_{ab} ([\g_\n, \g^\m])_c{}^d  \\
& \quad - \tfrac{1}{4} ([\g^\n, \g^\m])_{ab} (\g_\n)_c{}^d- 
\tfrac{1}{4}(\g^{5}[\g^\n, \g^\m])_{ab} (\g^{5}\g_\n
)_c{}^d\Big] \pa_\m\l_d \cr
~=&~ i 2 (\g^\m)_{ab} \pa_\m \l_c - Z^{(VS)}_{abc}  ~~,
\end{split}
\end{align}
where the on-shell central charge for the vector multiplet $Z^{(VS)}_{abc}$ given by
\begin{align}\label{e:ZVM}
\begin{split}
Z^{(VS)}_{abc}  ~=&~ i \tfrac{1}{4}\Big[ 2 (\g^\m)_{ab} \delta_c{}^d +(\g^\n
)_{ab} ([\g_\n, \g^\m])_c{}^d ~+~ ([\g^\n, \g^\m])_{ab} (
\g_\n)_c{}^d \\
& \quad {~~}+(\g^{5}[\g^\n, \g^\m])_{ab} (\g^{5}\g_\n
)_c{}^d\Big] \pa_\m \l_d   \cr
~=&~ i \tfrac{1}{4}\Big[2 (\g^\n)_{ab} (\g_\n \g^\m)_c{}^d 
~+~ ([\g^\n, \g^\m])_{ab} (\g_\n)_c{}^d \\
& \quad {~~}+(\g^{5}
[\g^\n, \g^\m])_{ab} (\g^{5}\g_\n)_c{}^d\Big]\pa_\m \l_d 
\\
~=&~ \Big[i \tfrac{1}{2}(\g^\n)_{ab} (\g_\n \g^\m)_c{}^d+ i\tfrac{1}{16} 
([\g_\a, \g_\b])_{ab}([\g^\a,\g^\b]\g^\m)_c{}^d  
\Big]\pa_\m \l_d  ~~.
\end{split}
\end{align}
In simplifying the central charge, we have used the identity in Eq.~(\ref{e:Fierz1}) as well 
as the following
\begin{align}\label{e:Fierz2}
([\g^\n, \g^\m])_{ab} (\g_\n)_c{}^d+(\g^{5}[\g^\n, \g^\m
])_{ab} (\g^{5}\g_\n)_c{}^d~=&~ \tfrac{1}{4} ([\g_\a, \g_\b])_{ab}(
[\g^\a,\g^\b]\g^\m)_c{}^d  ~~.
\end{align}
Notice the central charge $Z^{(VS)}_{abc}$ vanishes upon enforcing the equation of motion for 
the fermion $i(\g^\m)_c{}^d \pa_\m \l_d  = 0$ as it must\footnote{There is a typo 
in~\cite{Gates:2009me}, which disagrees in the last term of the last line of Eq.~(\ref{e:ZVM}) by a 
minus sign}.

\subsection{Axial-Vector Supermultiplet \texorpdfstring{$(U_{\m},\tilde{\l}_{c})$}{Um,lc}}
For the axial vector supermultiplet one obtains from simply repeating the steps used in (3.9) - 
(3.12) the results below.
\begin{align}
\{ \text{D}_a , \text{D}_b \} U_\m ~=&~ i 2 (\g^\n)_{ab} \pa_{\n} U_{\m} - \pa_\m 
\l^{(AVS)}_{ab}~~~,~~~\l^{(AVS)}_{ab} = i 2  (\g^\n)_{ab}U_\n ~~,
\end{align}
\begin{align}
\begin{split}
\{ \text{D}_a , \text{D}_b \} \tilde{\l}_c ~=&~  i 2 (\g^\m)_{ab} \pa_\m \tilde{\l
}_c - Z^{(AVS)}_{abc} ~~,
\end{split}
\end{align}
\begin{align}\label{e:ZAVM}
Z^{(AVS)}_{abc} ~&=~ \Big[i \tfrac{1}{2}(\g^\n)_{ab} (\g_\n \g^\m)_c{}^d-
 i\tfrac{1}{16} ([\g_\a, \g_\b])_{ab}([\g^\a,\g^\b]
 \g^\m)_c{}^d  \Big]\pa_\m \tilde{\l}_d  ~~.
\end{align}
Notice the slight difference in the central charges for the  vector multiplets and axial-vector 
multiplets, Eqs.~(\ref{e:ZVM}) and~(\ref{e:ZAVM}) respectively. The only difference is the 
minus sign of the second term. This is because the axial-vector and vector multiplets are 
related by the following replacements
\begin{align}
    A_\m \to U_\m~~~,~~~\l_a \to -i(\g^5)_a{}^b \tilde{\l}_b~~~.
\end{align}
Since $\g^5$ anti-commutes with all $\g^\m$, moving $\g^5$ past two $
\g^\m$ matrices in the leftmost term in the last line of Eq.~(\ref{e:ZVM}) gives two 
minus signs and moving it past three $\g^\m$ matrices in the rightmost term gives 
three minus signs.

\subsection{Matter-Gravitino Multiplet \texorpdfstring{$(B_{\m},\psi_{\m c})$}{Bm,pmc}}

The supermultiplet with propagating spins of one and three-halves has been called the 
``matter gravitino supermultiplet.''  Applying the procedure used in the previous sections
of this chapter yields the expressions in (3.17) - (3.24).
\begin{align}
\{ \text{D}_a , \text{D}_b \} B_\m ~=&~ i 2 (\g^\n)_{ab} \pa_{\n} B_{\m} - 
\pa_\m \l^{(MGM)}_{ab}~~~,~~~\l^{(MGM)}_{ab} = i 2  (\g^\n)_{ab}B_\n ~~,
\end{align}
\begin{align}
\begin{split}
\{ \text{D}_{a}, \text{D}_{b} \} \psi_{ \m c } ~=&~
i 2 (\g^\n)_{ab}\pa_\n \psi_{\m c} - \pa_\m \varepsilon^{(MGM)}_{abc} 
- Z_{\m abc}^{(MGM)}  ~~,  \cr
\end{split}
\end{align}
These include the gauge transformation
\begin{align}\label{e:epGaugeMGM}
\varepsilon^{(MGM)}_{abc} ~=&~ 2 i (\g^\a)_{ab}\psi_{\a c}     ~~,
\end{align}
for the spin 3/2 field and in addition the non-closure terms,
\begin{align}\label{e:ZMGM}
\begin{split}
Z_{\m abc}^{(MGM)} ~=&~
i (\g^\rho)_{ab} Z_{\m\rho c}^{(MGM,1)}  + i [\g^\rho , \g^\s]_{ab} 
Z_{\m\rho\s c}^{(MGM,2)}  ~~,
\end{split}
\end{align}
where
\begin{align}
Z_{\m\rho c}^{(MGM,1)} ~=&~ i\left[ \,   \left(-\eta_{\m\rho} + \tfrac{3}{4}\g_\m 
\g_\rho\right)_c{}^d   (\g^\k)_d{}^b 
- (\g_{[\m})_c{}^b \delta_{\rho]}{}^{\k}   \right] \,  E_{\k b} ~~, \\
Z_{\m\rho\s c}^{(MGM,2)} ~=&~ i\tfrac{1}{2}\left(\eta_{\m\rho}\delta_\s{
}^\a + i \tfrac{1}{2}\epsilon_{\m\rho\s}{}^\a \g^5 \right)_c{}^d 
\left( \delta_d{}^b  \delta_{\a}{}^{\k}  -  \tfrac{1}{4} (\g_\a 
\g^\k)_d{}^b  \right)\, E_{\k b}  ~~,
\end{align}
and $E_{\k b}$, 
\begin{align}\label{e:Epsi}
E_{\k b} ~=&~ \epsilon_\k{}^{\n\a\b} (\g^5 \g_\n)_b{
}^d\pa_\a \psi_{\b d}  ~~,
\end{align}
when set equal to zero, yields the Rarita-Schwinger equations of motion for the gravitino
and with its trace being
\begin{align}\label{e:traceEpsi}
R_a ~=&~ ([\g^\a, \g^\b])_a{}^b \pa_\a \psi_{\b b} = i
(\g^\k)_a{}^b E_{\k b} ~~~.
\end{align}
Notice the central charge terms vanish upon enforcement of the gravitino's 
equations of motion, i.e. setting $E_{\k b} = R_a = 0$, as they must.

\subsection{Supergravity Multiplet \texorpdfstring{$(h_{\m\n},\psi_{\m c})$}{hmn,pmc}}

For the supergravity supermultiplet the process goes along the now familiar
path and yields the results seen in (3.25) - (3.30).
\begin{align}
\{ \text{D}_{a}, \text{D}_{b} \} h_{\m\n}  ~=&~  i 2 (\g^{\rho})_{ab} \pa_{\rho} 
h_{\m\n} - \pa_{(\m}\xi_{\n)  ab}~~~,~~~\xi_{\n ab} =  i (\g^{\rho})_{ab} 
h_{\n\rho} ~~,
\end{align}
\begin{align}
\begin{split}
\{ \text{D}_{a}, \text{D}_{b} \} \psi_{ \m c } ~=&~ i 2 (\g^\n)_{ab}\pa_\n 
\psi_{\m c} - \pa_\m \varepsilon^{(SG)}_{abc} - Z_{\m abc}^{(SG)}  ~~,
\end{split}
\end{align}
\begin{align}\label{e:GaugeSG}
\varepsilon_{abc}^{(SG)} =~&  i \frac{1}{8} 
\left\{ 10 (\g^\a)_{ab} \delta_c{}^d  - [ \g^\a , \g^\b]_{ab} (
\g_\b)_c{}^d  + (\g_\b)_{ab} ([\g^\b , \g^\a])_c{}^d  
\right. \cr 
& \left. ~~~~ - (\g^5 [\g^\a , \g^\b])_{ab} (\g^5 
\g_\b)_c{}^d \right\}\psi_{\a d}  ~~,
\end{align}
\begin{align}\label{e:ZSG}
\begin{split}
Z_{\m abc}^{(SG)} ~=&~ i (\g^\rho)_{ab} Z_{\m\rho c}^{(SG,1)}  + i [\g^\rho , 
\g^\s]_{ab} Z_{\m\rho\s c}^{(SG,2)}  ~~,
\end{split}
\end{align}
where
\begin{align}
 Z_{\m\rho c}^{(SG,1)} =&  \tfrac{1}{4} ([\g_\m, \g_\rho])_c{}^d 
 R_d -i\tfrac{1}{2}   (\g_\m)_c{}^d E_{\rho d} +i\tfrac{3}{4} (\g_\rho
 )_c{}^d E_{\m d} ~~, \\
Z_{\m\rho\s c}^{(SG,2)} =& i\tfrac{1}{4}\eta_{\m\rho}E_{\s c} + 
i\tfrac{1}{32} ([\g_\rho, \g_\s])_c{}^d E_{\m d} + \tfrac{1}{8} 
\epsilon_{\m\rho\s}{}^\k (\g^5)_c{}^d E_{\k d} ~~,
\end{align}
with $R_a$ and $E_{\m a}$ taking the same form as for the matter gravitino supermultiplet.

With these results, we have reviewed what is known about the structure of on-shell
SUSY theories for fields of spins less than or equal to two.

\newpage
\section{A Geometry For Fermionic Non-closure Equation of Motion Terms}

In this chapter, we will focus on the equations of motion terms that generate central
charges in the SUSY algebra.

Let us start by gathering all of the results from the previous chapter in one place. 

\begin{align} \label{e:CMx}
Z^{(CS)}_{abc}&= i  (\g^\rho)_{ab}(\g_\rho\g^\m)_c{}^d \pa_\m \psi_d
\end{align}
\begin{align}\label{e:ZVMx}
\begin{split}
Z^{(VS)}_{abc}  
~=&~ i \Big[ \tfrac{1}{2}(\g^\n)_{ab} (\g_\n \g^\m)_c{}^d+ \tfrac{1}{16} 
([\g_\a, \g_\b])_{ab}([\g^\a,\g^\b]\g^\m)_c{}^d  
\Big]\pa_\m \l_d  ~~.
\end{split}
\end{align}
\begin{align}\label{e:ZAVMx}
Z^{(AVS)}_{abc} ~&=~  i \Big[ \tfrac{1}{2}(\g^\n)_{ab} (\g_\n \g^\m)_c{}^d-
\tfrac{1}{16} ([\g_\a, \g_\b])_{ab}([\g^\a,\g^\b]
 \g^\m)_c{}^d  \Big]\pa_\m \tilde{\l}_d  ~~.
\end{align}
\begin{align}\label{e:ZMGMx}
\begin{split}
Z_{\m abc}^{(MGM)} ~=&~
i (\g^\rho)_{ab} Z_{\m\rho c}^{(MGM,1)}  + i ([\g^\rho , \g^\s]){}_{ab} 
Z_{\m\rho\s c}^{(MGM,2)}  ~~,
\end{split}
\end{align}
where
\begin{align}
Z_{\m\rho c}^{(MGM,1)} ~=&~ i \left[ \,   \left(-\eta_{\m\rho} + \tfrac{3}{4}\g_\m 
\g_\rho\right)_c{}^d   (\g^\k)_d{}^b 
- (\g_{[\m})_c{}^b \delta_{\rho]}{}^{\k}   \right] \,  E_{\k b} ~~, \\
Z_{\m\rho\s c}^{(MGM,2)} ~=&~ i \tfrac{1}{2}\left(\eta_{\m\rho}\delta_\s{
}^\a + i \tfrac{1}{2}\epsilon_{\m\rho\s}{}^\a \g^5 \right)_c{}^d 
\left( \delta_d{}^b  \delta_{\a}{}^{\k}  -  \tfrac{1}{4} (\g_\a 
\g^\k)_d{}^b  \right)\, E_{\k b}  ~~,
\end{align}
with $E_{\k b}$
\begin{align}\label{e:Epsi2}
E_{\k b} ~=&~ \epsilon_\k{}^{\n\a\b} (\g^5 \g_\n)_b{
}^d\pa_\a \psi_{\b d}  ~~,
\end{align}
when set equal to zero, yields the Rarita-Schwinger equations of motion for the gravitino
and with its trace being
\begin{align}\label{e:traceEpsi2}
R_a ~=&~ ([\g^\a, \g^\b])_a{}^b \pa_\a \psi_{\b b} = 
i(\g^\k)_a{}^b E_{\k b} ~~~.
\end{align}
\begin{align}\label{e:GaugeSG2}
\varepsilon_{abc}^{(SG)} =~&  i \frac{1}{8} 
\left\{ 10 (\g^\a)_{ab} \delta_c{}^d  - [ \g^\a , \g^\b]_{ab} (
\g_\b)_c{}^d  + (\g_\b)_{ab} ([\g^\b , \g^\a])_c{}^d  
\right. \cr 
& \left. ~~~~ - (\g^5 [\g^\a , \g^\b])_{ab} (\g^5 
\g_\b)_c{}^d \right\}\psi_{\a d}  ~~,
\end{align}
\begin{align}\label{e:ZSG2}
\begin{split}
Z_{\m abc}^{(SG)} ~=&~ i (\g^\rho)_{ab} Z_{\m\rho c}^{(SG,1)}  + i [\g^\rho , 
\g^\s]_{ab} Z_{\m\rho\s c}^{(SG,2)}  ~~,
\end{split}
\end{align}
where
\begin{align}
 Z_{\m\rho c}^{(SG,1)} =&  \tfrac{1}{4} ([\g_\m, \g_\rho])_c{}^d 
 R_d -i\tfrac{1}{2}   (\g_\m)_c{}^d E_{\rho d} +i\tfrac{3}{4} (\g_\rho
 )_c{}^d E_{\m d} ~~, \\
Z_{\m\rho\s c}^{(SG,2)} =& i\tfrac{1}{4}\eta_{\m\rho}E_{\s c} + 
i\tfrac{1}{32} ([\g_\rho, \g_\s])_c{}^d E_{\m d} +\tfrac{1}{8} 
\epsilon_{\m\rho\s}{}^\k (\g^5)_c{}^d E_{\k d} ~~.
\end{align}

\newpage
\section{Review of 4D, \texorpdfstring{$\cal N$}{N} =1 Off-Shell Holoraumy Tensors From Previous Work}

In this chapter, we will present a review of off-shell holoraumy. 

For the spinor fields of the minimal representation supermultiplets (\ref{minSM1}) - 
 (\ref{minSM3}) we can introduce a ``collective'' notation whereby these spinors
 are denoted by the symbol $\Psi^{({\Hat {\cal R}})}_{a}$ where the ``representation
index $({\Hat {\cal R}})$'' distinguishes between the chiral/complex linear, vector,
or axial-vector supermultiplets.  Using this notation, the holoraumy calculations
take the form,
\be
 [\, \text{D}_{a} ~,~ \text{D}_{b} \, ]\, \Psi^{({\Hat {\cal R}})}_{c } ~=~ -\, i 2\,
 \big[ \hat{\bm h}^{\m} (\Pi) \big]_{abc}{}^{d} \,
 \pa_{\m} \Psi^{({\Hat {\cal R}})
 }_{d } ~~~,
\label{4dn2holor} \ee
where
\be  \eqalign{  {~~~}
 \big[ \hat{\bm h}^{\m} (\Pi) \big]_{abc}{}^{d} 
 &=~  {\big [} ~ {\rm p} \, C_{ab} \, (\g^{\m})_{c}{}^{d} ~+~ {\rm q}\,
(\g^{5})_{ab} \, (\g^{5}\g^{\m})_{c}{}^{d} \cr 
&{~~~~~~} ~~+~ {\rm r} \,  (\g^{5}\g^{\m})_{ab} \, (\g^{5})_{c}{}^{d} ~+~ \tfrac{1}{2} 
{\rm s} (\g^{5}\g_{\n})_{ab}\,  (\g^{5}[\g^{\n}, \g^{\m}])_{c}{}^{d} ~ {\big ]} 
~~~.
}   \label{Lilhs} \ee
and the integers p, q, r, and s take the values below\footnote{Here we follow the definitions of p, q, r, and s as in~\cite{4dHolor1,4dHolor2} as opposed to~\cite{4dHolor3} which used the opposite definitions of r and s.}.
\begin{table}[!h]
\centering
\begin{tabular}{c|cccc|}
$({\Hat {\cal R}})$  & ${\rm p}$ & ${\rm q}$ & ${\rm r}$ & ${\rm s}$  
\\ \hline 
(${\widehat {{CS}}}$) & 0 & 0 & 0 & 1    \\ 
(${\widehat {{VS}}}$) & 1 & 1 & 1 & 0    \\ 
(${\widehat {{AVS}}}$) & -1 & -1 & 1 & 0   \\ 
\end{tabular}
\caption{Holoraumy Integers For 4D, ${\cal N}=$ Chiral, Vector, and Axial-Vector
Supermultiplets.}
\label{tab:4DN2-pqrs}
\end{table} 

\noindent Finally, the quantity $\Pi$ is simply a notation for the set of integers p, q, r, and s
\begin{align}
	\Pi = (p,q,r,s)~~~.
\end{align}

While the anti-commutator algebra of the fermionic operator D${}_a$ with itself
takes a universal form being the sum of a translation operator, gauge transformation
operator on bosonic fields and including terms involving the equations of motion
for the fermionic fields, the commutator algebra of the fermionic operator D${}_a$
yields results that are specific to each off-shell supermultiplet.

As we are focused upon the on-shell holoraumy in this work, we will re-write (\ref{Lilhs})
to separate out the portions of those results that depend on the equations of motion for
the spinor in each supermultiplet from those that do not depend on such terms.  This
is done by ``splitting'' according
\be  \eqalign{  {~~~}
 \big[ \hat{\bm h}^{\m} (\Pi) \big]_{abc}{}^{d} 
 &=~  \big[ \hat{\bm h}^{\m\,EoM} (\Pi) \big]_{abc}{}^{d}
 ~+~ \big[ \hat{\bm h}^{\m\,On-Shell} (\Pi) \big]_{abc}{}^{d} \cr 
&=~  {\big [} ~ {\rm p} \, C_{ab} \, (\g^{\m})_{c}{}^{d} ~+~ {\rm q}\,
(\g^{5})_{ab} \, (\g^{5}\g^{\m})_{c}{}^{d} ~+~ 
{\rm s} (\g^{5}\g_{\n})_{ab}\,  (\g^{5} \g^{\n} \g^{\m})_{c}{}^{d}  ~ {\big ]}  \cr 
&{~~~~~~} ~~+~ 
(\, {\rm r} \, -\,  {\rm s}  \, ) \,(\g^{5}\g^{\m})_{ab} \, (\g^{5})_{c}{}^{d} 
~~~,
}   \label{Lilhs2} \ee
where all we have done is to make use of the identity
\be
 (\g^{5}[\g^{\n}, \g^{\m}])_{c}{}^{d} ~=~ 
  (\g^{5}\g^{\n} \g^{\m})_{c}{}^{d}  ~-~ (\g^{5}\g^{\m} \g^{\n})_{c}{}^{d}
  ~=~ - \, 2  \eta{}^{\m \n} (\g^{5})_{c}{}^{d} ~+~ 2\,
  (\g^{5}\g^{\n} \g^{\m})_{c}{}^{d}  ~~~,
\ee
which is valid for the $\g$-matrices.  

Accordingly, the first line of terms on the RHS of (\ref{Lilhs2}) are all proportional 
to the equations of motion for the fermions, while the single term on the second 
line is not. Owing to the Clifford algebra satisfied by the $\gamma^\mu$-matrices, 
Eqs.~(\ref{Lilhs}) and~(\ref{Lilhs2}) are just two of an infinite number of ways to 
separate $ \hat{\bm h}^{\m} (\Pi)$ in terms proportional to equations of motion 
and terms that are not. Our convention will be to simply commute the gamma 
matrix that contracts with the derivative (and/or gravitino for the $SG$ and $MGM$ 
multiplets we will investigate) to the right until it is no longer possible to produce 
terms proportional to equations of motion by doing so.

The equation in (\ref{Lilhs2}) reveals that all the information associated with p, q, and s 
separately is lost (as these occur in the $ \big[ \hat{\bm h}^{\m\,EoM} (\Pi) \big]$ tensor) 
for the on-shell theories describing the minimal representations. We have shown by
first calculating the off-shell fermionic holoraumy and then eliminating the portion of it that
depends on the equations of motion for the fermions leads to
\be
 [\, \text{D}_{a} ~,~ \text{D}_{b} \, ]\, \Psi^{({\Hat {\cal R}})}_{c } ~=~ -\, i 2\,
 (\, {\rm r} \, -\,  {\rm s}  \, ) \,(\g^{5}\g^{\m})_{ab} \, (\g^{5})_{c}{}^{d}  \,
 \pa_{\m} \Psi^{({\Hat {\cal R}})
 }_{d } ~~~,
\label{4dn2holorOS} \ee
So the only information 
retained relates to the combination (r $-$ s) which occurs in the $ \big[ \hat{\bm h}^{\m
\,On-Shell} (\Pi) \big]$ tensor.  Thus, there is a severe loss of information upon enforcing the equations of motion.  We return to 
this in the conclusions.

\newpage
\section{New Results of 4D, \texorpdfstring{$\cal N$}{N} =1 On-Shell Holoraumy Tensors}

The general method to calculate the fermionic holoraumy is an expansion in a
basis of antisymmetric matrices
\begin{equation}
[\, \text{D}_{a} ~,~ \text{D}_{b} \, ] ~=~ \fracm{1}{2}\, C_{ab} \,  C^{cd}  
 \text{D}_{c} \text{D}_{d}
+ \fracm{1}{2} \,  (\g^{5})_{ab}  (\g^{5})^{cd}  \text{D}_{c} \text{D}_{d} 
- \fracm{1}{2} \,  (\g^{5} \g^{\k})_{ab}  (\g^{5}
\g_{\k} )^{cd}  \text{D}_{c} \text{D}_{d} 
\end{equation}
due to the identities
\begin{equation}
\begin{aligned}
 C^{ab}C_{ab} ~=& ~4  ~~, &
 C^{ab}(\g^{5})_{ab} ~=& ~0 ~~,  &
 C^{ab}(\g^{5}\g^{\k})_{ab} ~=& ~0 ~~,   \\
 & &  (\g^{5})^{ab}(\g^{5})_{ab} ~=& ~4 ~~,  &
 (\g^{5})^{ab}(\g^{5}\g^{\k})_{ab} ~=& ~0 ~~,   \\
 & &  & &  (\g^{5}\g^{\rho})^{ab}(\g^{5}\g^{\k})_{ab} ~=& ~- 
 4 \eta^{\rho\k} ~~.
\end{aligned}
\end{equation}
So all that remains is to evaluate the terms quadratic in the D-operators on the RHS 
of (\ref{Lilhs2}) making use of the expressions in (\ref{minSM1}) -  (\ref{minSM5}). 

\subsection{Chiral/Complex Linear Supermultiplet On-Shell 
\texorpdfstring{$(A,B,\psi_{c})$}{A,B,pc}}

We begin with the spin-0 bosons and derive
\begin{align}
	[\text{D}{}_a , \text{D}{}_b] A =& -2 (\g^5\g^\m)_{ab} \pa_\m B  ~~,  \\
    [\text{D}{}_a , \text{D}{}_b] B =& 2 (\g^5\g^\m)_{ab} \pa_\m A  ~~~~~~,
\end{align}
and find for the fermion
\begin{align}
[\text{D}{}_a , \text{D}{}_b] \psi_c = i(\g^5\g^\m)_{ab}(\g^5)_c
{}^d\pa_\m \psi_d - i \tfrac{1}{2} (\g^5\g^\k)_{ab}(\g^5
[\g_\k,\g^\m]){}_c{}^d\pa_\m \psi_d  ~~.
\end{align}
These can be cast into the following forms
\begin{align}
[\text{D}{}_a , \text{D}{}_b] \begin{pmatrix} A \\  B \end{pmatrix} 
=& -i 2 \mathscr{B}_{ab}^{\m (CS)}\pa_\m \begin{pmatrix} A \\  B \end{pmatrix}   
~~, 
\label{axRot0b}
\end{align}
for the bosonic fields and for the fermions as
\begin{align}
[\text{D}{}_a , \text{D}{}_b] \psi_c =& - i 2  \left(\mathscr{F}^{\m(CS)}_{ab}\right)_c{
}^{d}\pa_\m \psi_d  - \mathscr{Z}_{abc}^{(CS)}
~~,
\label{axRot0f}
\end{align}
where the on-shell bosonic and fermionic holoraumy tensors $\mathscr{B}_{ab}^{\m 
(CS)}$, and $\left(\mathscr{F}^{\m(CS)}_{ab}\right)_c{}^{d}$, along with the EoM 
portion 
$\mathscr{Z}_{abc}^{(CS)}$ are explicitly written as
\begin{align}
\mathscr{B}_{ab}^{\m (CS)} =& \, \, \s^2 \otimes (\g^5\g^\m)_{ab}   ~~,
\end{align}
\begin{align}
\left(\mathscr{F}_{ab}^{\m (CS)}\right)_c{}^d &=   -(\g^5\g^\m)_{ab}(\g^5)_c{}^d  ~~,
\label{fHL}
\end{align}
\begin{align}
\mathscr{Z}_{abc}^{(CS)} = & ~  (\g^5\g^\k)_{ab}(\g^5\g_\k)_c{}^d  
\mathcal{D}_d^{(CS)} ~~,
\end{align}
where
\begin{align}
\mathcal{D}_d^{(CS)} =& i (\g^\m)_d{}^e \pa_\m \psi_e  ~~.
\end{align}
and when this is set to zero it yields the equation of motion for $\psi_e$, thus it
is acting like a central charge for holoraumy.

These results also reveal something else...the presence of an electromagnetic 
type duality rotation.  This is seen in the following manner.

In electromagnetism, if we start with the source-free Maxwell Equations, they 
are invariant under a transformation where we `rotate' the electric field ${\vec E}$
and magnetic field ${\vec B}$ one into the other via
\be  \eqalign{
{\vec E}' ~&=~ {\vec E} \, cos \Theta ~-~ {\vec B} \, sin \Theta ~~~, \cr
{\vec B}' ~&=~  {\vec E} \, sin \Theta ~+~ {\vec B} \, cos \Theta ~~~.
} 
\label{axRot}
\ee
Since ${\vec E}$ transforms as a vector under a parity transformation of field 
variables while ${\vec B}$ transforms as an axial-vector under the same 
transformation, the rotation in (\ref{axRot}) mixes field variables that possess 
different parity properties.  Now let us note that bosonic field variables ($A$,
$B$) of the chiral supermultiplet are just such a pair,  thus under the transformation
of (\ref{axRot}), the electric and magnetic field variable can be replaced by
the  of the derivatives of the spin-0 scalar and pseudoscalar field variables
leading to
\be  \eqalign{
{(\pa{}_{\m} A)}' ~&=~ {(\pa{}_{\m} A)} \, cos \Theta ~-~ {(\pa{}_{\m} B)} \, sin \Theta ~~~, \cr
{(\pa{}_{\m} B)}' ~&=~  {(\pa{}_{\m} A)} \, sin \Theta ~+~ {(\pa{}_{\m} B)} \, cos \Theta ~~~.
} \label{axRot2}
\ee
Upon comparing the form of the RHS of (\ref{axRot0b}) with (\ref{axRot2}),
we see the angle variable has the value of  $\Theta$ = $\pi/2$.

This electromagnetic duality rotation must also effect the $\g$-matrices since
$\g{}^{\m}$ and $i\g{}^5\g{}^{\m}$ must satisfy a condition similar in form to
the equations in (\ref{axRot}) and (\ref{axRot2}).
\be  \eqalign{
({\g{}^{\m}})' ~&=~ {\g{}^{\m}} \, cos \Theta ~-~ {i\g{}^5\g{}^{\m}} \, sin \Theta ~~~, \cr
({i\g{}^5\g{}^{\m}})' ~&=~  {\g{}^{\m}} \, sin \Theta ~+~ {i\g{}^5\g{}^{\m}} \, cos \Theta ~~~.
} 
\label{axRot3}
\ee
The 2 $\times$ 2 matrix that appears in (\ref{axRot}) - (\ref{axRot3}) for 
convenience be written as
\begin{equation}
 {\cal P}(\Theta)~=~
\begin{bmatrix}
cos \Theta & - sin \Theta \\
sin \Theta & cos \Theta \\
\end{bmatrix} 
\end{equation}

As a consistency check, we need to consider the on-shell fermionic holoraumy.
i.\ e.\ the first term on the RHS of ({\ref{axRot0f}).  The presence of the factor
of $\g^5 \g{}^{\m}$ in ({\ref{axRot0f}) show the usual $\g{}^{\m}$ associated
with the SUSY algebra has been subjected to electromagnetic duality rotation
where $\Theta$ = $\pi/2$.  However, the complete form of $(\mathscr{F}_{ab}^{
\m (CS)}){}_c{}^d$ shows the electromagnetic duality rotation is realized as
an operator 
\be  \eqalign{
(\psi{}_a)' ~&=~  cos \Theta \,   \psi{}_a
~-~ i \, sin \Theta \, \,  (\g^{5}){}_a{}^b \psi{}_b
  ~~~.
} 
\label{axRot4}
\ee
acting on the on-shell fermion as well with the same value of $\Theta$.

In fact, in the remainder of this section, we find these conditions are satisfied
uniformly on all of the supermultiplets under explicit investigation.

\subsection{Vector Supermultiplet \texorpdfstring{$(A_{\m},\l_{c})$}{Am,lc}}

The on-shell holoraumy calculations for the vector supermultiplet yield
\begin{align}
[\text{D}{}_a,\text{D}{}_b] A_\m = 
- i 2 \left(\mathscr{B}_{\m\a\b}^{(VS)} \right)_{ab}	\pa^\a A^\b
~~~,
\label{ph1}
\end{align}
\begin{align}
\begin{split}
[\text{D}{}_a,\text{D}{}_b] \l_c 
=& - i 2  \left(\mathscr{F}^{\m(VS)}_{ab}\right)_c{}^{d}\pa_\m \l_d 
- \mathscr{Z}_{abc}^{(VS)} ~~~,
\end{split}
\label{ph2}
\end{align}
with the expressions for the bosonic and fermionic holoraumy tensors given 
by
\begin{align}\label{e:BVM}
\left(\mathscr{B}_{\m\a\b}^{(VS)} \right)_{ab} =& - i \epsilon_{\m\a
\beta\n}(\g^5\g^\n)_{ab} ~~, \\
\label{e:FVM}
\begin{split}
\left(\mathscr{F}^{\m(VS)}_{ab}\right)_c{}^{d} =& (\g^5\g^\m)_{ab} (\g^5)_c{}^d
~~,
\end{split}
\end{align}
along with the fermionic equation of motion term
\begin{align}
\mathscr{Z}_{abc}^{(VS)} =&  \Big\{\tfrac{3}{2}  C_{ab}\delta_c{}^d+\tfrac{3}{2}  
(\g^5)_{ab}(\g^5)_c{}^d - \tfrac{1}{2} (\g^5\g^\k)_{ab}(\g^5\g_\k)_c{}^d 
\Big\} \mathcal{D}_d^{(VS)} ~~,
\end{align}
with
\begin{align}
\mathcal{D}_d^{(VS)} =& i (\g^\m)_d{}^e \pa_\m \l_e ~~.
\end{align}
The term 
$\mathscr{Z}_{abc}^{(VS)}$ vanishes upon enforcing the equations of motion 
$\mathcal{D}_d^{(VS)} = 0$.

To check for the presence of an electromagnetic duality rotation as found for
the case of the on-shell chiral/complex linear supermultiplet has two parts,
i.\ e.\ an examination of the bosonic holoraumy and an examination of the 
fermionic holoraumy. 

For the gauge field, we see combine the result in (\ref{ph1}) with the one
in (\ref{e:BVM}) to write
\begin{align}
[\text{D}{}_a,\text{D}{}_b] A_\m = 
-  \epsilon_{\m\a
\beta\n}(\g^5\g^\n)_{ab} \,  F{}^{\a  \b}(A) ~=~ - 2 
(\g^5\g^\n)_{ab} \,  {\Tilde F}{}_{\m  \n}(A)
~~~,
\label{ph3}
\end{align}
where in the middle step the Maxwell Field Strength $F{}^{\a  \b}(A)$ is 
introduced and in the final step the answer is expressed in terms of the 
dual Maxwell Field Strength ${\Tilde F}{}_{\m \n}(A)$ = $\fracm 12 \e_{\m \n 
 \a \b}F{}^{\a  \b}(A)$.  Of course, (\ref{ph3}) is simply the manifestly 
 relativistic formulation of (\ref{axRot}).
 
Upon comparing this result with the one 
 in (\ref{susyMax}), it clear that an electromagnetic duality rotation in present.
It is realized precisely as in (\ref{axRot}) though the same angle
$\Theta$ = $\pi$/2.

The explicit introduction of the dual electromagnetic field strength also
allows the discussion of the transformation under the action of the 
electromagnetic duality rotation according to
\be  \eqalign{
{ F{}_{\a  \b}(A)}' ~&=~ { F{}_{\a  \b}(A)} \, cos \Theta ~-~ {{\Tilde F}{}_{\a 
 \b}(A)} \, sin \Theta ~~~, \cr
{{\Tilde F}{}_{\a  \b}(A)}' ~&=~  { F{}_{\a  \b}(A)} \, sin \Theta ~+~ {{\Tilde 
F}{}_{\a  \b}(A)} \, cos \Theta ~~~.
} 
\label{axRotF}
\ee
Before leaving this point, it is also important to keep in mind that the
electromagnetic duality rotation defined by the formula (\ref{axRotF})
can be applied to any field that possesses a field strength that is a
two-form.  This includes spin-3/2 fields as well as spin-2 fields where
the first two (in our conventions) indices on the Riemann curvature
tensor are treated as the indices on the Maxwell field strength.

For the gaugino field, we may compare the result for the fermionic
holoraumy in (\ref{fHL}) for the spinor in the chiral/complex linear
supermultiplet with the one in (\ref{e:FVM}) for the vector supermultiplet.
There is a sign difference seen upon doing so.  This is a feature to be
discussed in the conclusion section.

\subsection{Axial-Vector Supermultiplet}

The calculations for the axial-vector supermultiplet simply reproduce
exactly the ones carried out upon the vector supermultiplet.  So in the
discussion below we simply report the results for the sake of explicitness,
but without further comment. 
\begin{align}
[\text{D}{}_a,\text{D}{}_b] U_\m =&
- i 2 \left(\mathscr{B}_{\m\a\b}^{(AVS)} \right)_{ab}	
\pa^\a U^\b
\end{align}

\begin{align}
\begin{split}
[\text{D}{}_a,\text{D}{}_b] \tilde{\l}_c =& - i 2  \left(\mathscr{F}^{\m
(AVS)}_{ab}\right)_c{}^{d}\pa_\m \tilde{\l}_d- \mathscr{Z}_{
abc}^{(AVS)}
\end{split}
\end{align}
where the bosonic and fermionic holoraumy tensors are
\begin{align}\label{e:BAVM}
\left(\mathscr{B}_{\m\a\b}^{(AVS)} \right)_{ab} =& - i \epsilon_{
\m\a\b\n}(\g^5\g^\n)_{ab} \\
\label{e:FAVM}
\left(\mathscr{F}^{\m(AVS)}_{ab}\right)_c{}^{d} =&  (\g^5\g^\m)_{ab
} (\g^5)_c{}^d \\
\label{E:ZAVM}
\mathscr{Z}_{abc}^{(AVS)} =& \Big\{ -\tfrac{3}{2}C_{ab}\delta_c{}^d
- \tfrac{3}{2}(\g^5)_{ab}(\g^5)_c{}^d- \tfrac{1}{2} (\g^5\g^\k)_{ab}
(\g^5\g_\k)_c{}^d\Big\} \mathcal{D}_d^{(AVS)}
\end{align}
where
\begin{align}
\mathcal{D}_d^{(AVS)} =& i(\g^\m)_d{}^e \pa_\m \tilde{\l}_e
\end{align}
is when set to zero the equation of motion for $\tilde{\l}_e$.  Notice 
the term $\mathscr{Z}_{abc}^{(AVS)}$ vanishes upon enforcing the equations 
of motion $\mathcal{D}_d^{(AVS)} = 0$.

As there has not been any published work looking at the off-shell holoraumy
of the matter-gravitino supermultiplets, we will not in the following to be making 
comparisons between on-shell and off-shell holoraumies.  However, as there
are spin-1 fields in the vector, axial vector, and matter gravitino supermultiplets,
we will make comments on this at the end of the next subsection.

\subsection{Matter-Gravitino Multiplets \texorpdfstring{$(B_{\m},\psi_{\m c})$}{Bm,pmc}}

\begin{align}
[\text{D}{}_a,\text{D}{}_b] B_{\m} =&	-i 2 \left(\mathscr{B}_{\m\a\b}^{(MGM)} 
\right)_{ab}	\pa^\a B^\b \\
\label{e:MGMFHolo}
[\text{D}{}_a, \text{D}{}_b] \psi_{\m c} =& - i 2  \left(\mathscr{F}_{\m\a\b}^{(MGM)} 
\right)_{abc}{}^d \pa^\a \psi^{\b}{}_{d} - \mathscr{Z}_{\m abc}^{(MGM)} .
\end{align}

where the bosonic and fermionic holoraumy tensors are
\begin{align}\label{e:BMGM}
\left(\mathscr{B}_{\m\a\b}^{(MGM)} \right)_{ab} =& i \epsilon_{\m\a
\b\n}(\g^5\g^\n)_{ab} \\ 
\label{e:FMGM}
\begin{split}
\left(\mathscr{F}_{\m\a\b}^{(MGM)} \right)_{abc}{}^d =&    \tfrac{1}{2} \eta_{\m[\a}\eta_{\b]\n}  (\gamma^{5}\gamma^{\n})_{ab} (\gamma^{5})_{c}{}^{d}   - i\tfrac{1}{2} \epsilon_{\m\a\b\n}  (\gamma^{5}\gamma^{\n})_{ab} \delta_{c}{}^{d} 
\end{split} \\
\label{e:ZHoloMGM}
\begin{split}
	\mathscr{Z}_{\m abc}^{(MGM)} =&  i \tfrac{1}{4}  \Big\{  C_{ab} (\gamma_\m)_{c}{}^{d} -  (\gamma^{5})_{ab} (\gamma^{5}\gamma_\m)_{c}{}^{d}   - (\gamma^{5}\gamma^{\rho})_{ab} (\gamma^{5}\gamma_{\rho}\gamma_\m)_{c}{}^{d}  \Big\}   R_d\\
    &  +\Big\{  C_{ab} \delta_{c}{}^{d}  -  (\gamma^{5})_{ab} (\gamma^5)_{c}{}^{d}     \Big\}  E_{\m d}
\end{split}
\end{align} 
where $E_{\m d}$ and $R_a$ are as in Eqs.~(\ref{e:Epsi}) and~(\ref{e:traceEpsi}), 
when set to zero the equation of motion for the gravitino and its trace, respectively. 
We see then that $\mathscr{Z}_{\m a b c}^{(MGM)}$ vanishes upon enforcing the 
equations of motion for the gravitino.

Looking at the result in (\ref{e:BVM}) in comparison to the one in (\ref{e:BMGM}),
it can be seen that
\be
\left(\mathscr{B}_{\m\a\b}^{(VS)} \right)_{ab} ~=~ - \left(\mathscr{B}_{\m\a\b}^{
(MGM)} \right)_{ab} 
\ee
which is one more of the points to be covered in the concluding discussion

\subsection{Supergravity Multiplets \texorpdfstring{$(h_{\m\n},\psi_{\m c})$}{
hmn,pmc}}

Finally, we come to the case of the on-shell supergravity supermultiplets whose
results are reported in this subsection.

\begin{align}
[\text{D}{}_a,\text{D}{}_b] h_{\m\n}  =& - i 2 \left(\mathscr{B}_{\m\n\rho\a
\b}^{(SG)} \right)_{ab}	\pa^\rho h^{\a\b} \\
[\text{D}{}_a, \text{D}{}_b] \psi_{\m c} =& - i 2 \left(\mathscr{F}_{\m\a\b}^{
(SG)} \right)_{abc}{}^d \pa^\a \psi^\b{}_d - \mathscr{Z}_{\m abc}^{(SG)} - 
\pa_\m \zeta_{abc}^{(SG)}
\end{align}
where the bosonic and fermionic holoraumy tensors are
\begin{align}\label{e:BSG}
\left(\mathscr{B}_{\m\n\rho\a\b}^{(SG)} \right)_{ab} =&  -i \tfrac{1}{2} \eta_{
\b (\m }\epsilon_{\n)\rho\a\k}  ( \g ^{5} \g^{\k} )_{ab}  
\end{align}
\begin{align}
\label{e:FSG}
\left(\mathscr{F}_{\m\a\b}^{(SG)} \right)_{abc}{}^d =&  -\tfrac{3}{4} \eta_{
\m[\a} (\g^5\g_{\b]})_{ab}(\g^5)_c{}^d + i \tfrac{1}{8} \epsilon_{\m\a
\b\n} (\g^5\g^\n)_{ab}\delta_c{}^d + \tfrac{1}{16} (\g^5\g_{[\a})_{ab}(\g^5
[\g_{\b]}, \g_\m])_c{}^d \\
\label{e:ZHoloSG}
\begin{split}
\mathscr{Z}_{\m abc}^{(SG)} = &i\tfrac{1}{2}\Big\{ - C_{ab} (\g_\m)_c{}^d - (\g^5
)_{ab}(\g^5\g_\m)_c{}^d+ \tfrac{1}{2} (\g^5\g^\s)_{ab}(\g^5\g_\s\g_\m
)_c{}^d  \cr
&\hspace{30pt} -\tfrac{1}{4} (\g^5\g_\m)_{ab} (\g^5)_c{}^d\Big\}R_d \cr
& - \tfrac{1}{4}  \Big\{5 C_{ab}\delta_c{}^d + 5  (\g^5)_{ab}(\g^5)_c{}^d - 
2  (\g^5\g^\s)_{ab}(\g^5\g_\s)_c{}^d\Big\}E_{\m d}
\end{split} \\
\label{e:GaugeHoloSG}
\zeta_{abc}^{(SG)} =& i\tfrac{3}{4} \Big\{ C_{ab} (\g^\b)_c{}^d + (\g^5)_{ab}
(\g^5\g^\b)_c{}^d + (\g^5\g^\b)_{ab}(\g^5)_c{}^d  \cr
&\hspace{30pt}- \tfrac{1}{6} (\g^5 \g_\k)_{ab}(\g^5[\g^\k , 
\g^\b])_c{}^d \Big\} \psi_{\b d}
\end{align}
Notice all terms  in $\left(\mathscr{F}_{\m\a\b}^{(SG)}\right)_{abc}{}^d$ are 
anti-symmetric in $\a$ and $\b$ as in the case for $MGM$. Also, we see that 
$\mathscr{Z}_{\m abc}^{(SG)}$ vanishes upon enforcing the equations of motion 
for the gravitino and its trace, Eqs.~(\ref{e:Epsi}) and~(\ref{e:traceEpsi}) set to zero, 
respectively. The term $\zeta_{abc}^{(SG)}$ acts like a gauge transformation that 
can not be included in  $\left(\mathscr{F}_{\m\a\b}^{(SG)}\right)_{abc}{}^d$ 
while maintaining the $\a$ and $\b$ antisymmetry.  Comparing the $SG$ 
fermionic holoraumy to that of $MGM$, Eq. (\ref{e:MGMFHolo}), we point out that 
the former has a gauge-like term $\zeta_{abc}^{(SG)}$ that remains upon enforcing 
the equations of motion and the latter does not.

\newpage
\section{Conclusion}

Perhaps, one of the fascinating observations uncovered by this study in the fact that
the commutator, not the anti-commutator of two supercharges in the theories under
investigation show a previously unnoticed uniformity.

Putting all of this together, we arrive at a conclusion that to our knowledge has
not appeared in the physics literature.  In a field independent manner, for an
on-shell supermultiplet in four dimensions upon dropping dependences proportional
to equations of motion and gauge transformations, the equation 
\be
[\, \text{D}{}_a ~,~ \text{D}{}_b \, ] ~=~  2 \, {\cal Q}_{EMDC} \,
{\cal P}(\Theta = \pi/2) \, \otimes \,
(\g^\m)_{ab}  {\cal P}{}_{\m}  ~~~,
\label{axROT}
\ee
with \newline \indent
(a.) $ {\cal P}(\Theta = \pi/2)$ being an electromagnetic duality 
      rotation through \newline \indent 
      $~~~~~$ an angle of $\pi/2$ acting on all quantities 
      that follow it,
 \newline \indent     
(b.) ${\cal P}{}_{\m}$ being 
the generator of spacetime translations is satisfied, and
 \newline \indent   
 (c.)  ${\cal Q}_{EMDC} $ denoting an electromagnetic-duality charge,
\vskip.1in \noindent
appears to be ubiquitously valid.

As several points in the discussions of the holoraumy tensors, we alluded to
the fact that the signs that show up in some of the calculation might seem inconsistent
with the action of the operator $ {\cal P}(\Theta = \pi/2)$ realized on the fields of
the various multiplets.  The simplest way to resolve this is to conclude that when
action on field variable, the operator $ {\cal P}(\Theta = \pi/2)$ is also dependent
on ``an electromagnetic-duality charge" and different spinors in the distinct supermultiplets
carry different values of this charge.

Let us give an expanded discussion of this ``electromagnetic-duality charge" that can be
denoted by ${\cal Q}_{EMDC}$.

The calculations reviewed in chapter four are off-shell calculation where sets
of auxiliary fields are included ab initio.  We used the form of the off-shell holoraumy
to separate it into on part proportional to terms involving equations of motion
(the first line shown in (\ref{Lilhs2}) and a single term not proportional to 
the equations of motion (i. e. the fine line of (\ref{Lilhs2}).  Thus a single
term for the on-shell holoraumy, shown in (\ref{4dn2holorOS}), remained. 
The implication of (\ref{4dn2holorOS}) is
\be
{\cal Q}_{EMDC} ~=~  (\, {\rm r} \, -\,  {\rm s}  \, )  ~~~.
\ee
With this understanding in place, all the signs are consistent and all the calculations
take the form of the equation in (\ref{axROT}).

We conjecture this is valid for {\em {all}} on-shell 4D, $\cal N$ = 1 supermultiplets.

Of course, we always have the usual condition
\be
\{\, \text{D}{}_a ~,~ \text{D}{}_b \, \} ~=~  2 \, 
(\g^\m)_{ab}  {\cal P}{}_{\m}  ~~~,
\label{susy}
\ee
and upon adding (\ref{axROT}) and (\ref{susy}) we obtain the very powerful
statement (without commutators nor anti-commutators of the supercovariant
derivatives)
\be
\text{D}{}_a \, \text{D}{}_b \,  ~=~  (\g^\m)_{ab}  {\cal P}{}_{\m}
~+~ {\cal Q}_{EMDC} \, {\cal P}(\Theta = \pi/2) \, \otimes \,
(\g^\m)_{ab}  {\cal P}{}_{\m}  ~~,
\ee
whose implications need further study.

We will close by carrying out a simple test of this construction in (\ref{axROT}) as this
equation makes a prediction that is simple to verify.  In chapter two, there were two
minimal 4D, $\cal N$ = 1 supermultiplets that were not included...the tensor supermultiplet
and the axial tensor supermultiplet.  The reason for neglecting them is that the form
of their supersymmetry variations of the fields $(\varphi, \, B{}_{\mu \, \nu }, \,  \chi_a )$
\be
 \eqalign{
{\rm D}_a \varphi ~&=~ \chi_a  ~~~,~~~~~ 
{\rm D}_a B{}_{\mu \, \nu } ~=~ -\, \fracm 14 ( [\, \gamma_{\mu}
\, , \,  \gamma_{\nu} \,]){}_a{}^b \, \chi_b  ~~~, \cr
{\rm D}_a \chi_b ~&=~ i\, (\gamma^\mu){}_{a \,b}\,  \partial_\mu \varphi 
~-~  (\gamma^5\gamma^\mu){}_{a \,b} \, \e{}_{\mu}{}^{\r \, \s \, \t}
\partial_\r B {}_{\s \, \t}~~, {~~~~~~~~~~~~~~\,~~}
}  \label{QT3}
\ee
are seen to be the same whether on-shell or off-shell.  On the basis of (\ref{axROT}) and
{\em {without}} a single calculation one can predict the on-shell bosonic holoraumy 
must involve an electromagnetic duality rotation being realized as
\be \eqalign{
({\pa_{\m} \varphi})' ~&=~ ({\pa_{\m} \varphi}) \, cos \Theta ~-~   \epsilon_{
\m \k \l \n}( \pa{}^{\k} B{}^{\l \n} )  \, sin \Theta ~~~, \cr
\epsilon_{\m \k \l \n}( \pa{}^{\k} 
B{}^{\l \n} ) ' ~&=~  ({\pa_{\m} \varphi}) \, sin \Theta ~+~ \epsilon_{\m \k \l \n}
( \pa{}^{\k} B{}^{\l \n} )  \, cos \Theta ~~~.
} 
\label{axRotF2}
\ee
The precise results of the calculation are
\begin{align}\label{e:HoloPhiTensor}
	[\rD_a , \rD_b] \varphi =& - 2 (\gamma^5 \gamma^\mu)_{ab} \epsilon_\mu{}^{\rho\sigma\tau} \partial_\rho B_{\sigma\tau} \\
	\label{e:HoloBTensor}
	[\rD_a , \rD_b] B_{\mu\nu} =& \epsilon_{\mu\nu\alpha\beta}(\gamma^5\gamma^\alpha)_{ab}\partial^\beta \varphi + (\gamma^5 \gamma_{[\mu})_{ab}\epsilon_{\nu]\rho\sigma\tau}\partial^\rho B^{\sigma\tau}\\
	\label{e:HolochiTensor}
	[\rD_a , \rD_b]  \chi_c =& - 2 i \left(\mathscr{F}^{\mu(TS)}_{ab}\right)_{c}{}^d \partial_\mu \chi_d - \mathscr{Z}^{(TS)}_{abc}
\end{align}
where
\begin{align}\label{e:FHoloTensor}
\begin{split}
	\left(\mathscr{F}^{\mu(TS)}_{ab}\right)_{c}{}^d =& - (\gamma^5\gamma^\mu)_{ab} (\gamma^5)_c{}^d ~~~,
	\end{split} \\
	\label{e:ZHoloTensor}
	\mathscr{Z}^{(TS)}_{abc} =& 2 \left\{- \,C_{ab}\delta_c{}^d +  (\gamma^5)_{ab}(\gamma^5)_c{}^d  \right\} \mathcal{D}_d^{(TS)}\\
	\label{e:DTS}
	\mathcal{D}^{(TS)}_d =& i (\gamma^\mu)_{d}{}^e\partial_\mu \chi_e
\end{align}
Neglecting the second term in the holoraumy of $B_{\mu\nu}$ and comparing Eqs.~(\ref{e:HoloPhiTensor}) and~(\ref{e:HoloBTensor}) to~(\ref{axRotF2}) demonstrates that $\Theta = \pi/2$ as above. 
Thus, in the end, the appearance of the electromagnetic duality rotation within the holoraumy
in on-shell 4D,  $\cal N$ = 1 is the final result of parity conservation.

One final observation about the presences of these electromagnetic duality rotations is the
role they play in describing the dynamics of 4D, $\cal N$ = 1 supersymmetrical systems.
The evidence from our investigation here is that these electromagnetic duality rotations
are ubiquitously present in all 4D, $\cal N$ = 1 supersymmetrical theories.

There are three parts to understanding why this must be so.

(a.)
As the example of the tensor supermultiplet shows, the off-shell holoraumy
\newline $~~~~~~~~~~~$ tensors can 
always be decomposed into one portion that includes the duality
\newline $~~~~~~~~~~~$ rotations
and then other terms.  Looking closely at the terms in 
(\ref{e:HoloPhiTensor}) and
\newline $~~~~~~~~~~~$ 
(\ref{e:HoloBTensor}) on the bosonic fields in the supermultiplet, while
the leading term in
\newline $~~~~~~~~~~~$
(\ref{e:HolochiTensor}) shows show the effect of the electromagnetic 
duality rotations on the
\newline $~~~~~~~~~~~$
fermionic term.

(b.)
This general behavior can be seen in all off-shell holoraumy calculations
\newline $~~~~~~~~~~~$
in 4D, $\cal N$ = 1 supermultiplets~\cite{4dHolor2}.

(c.)
The final portion to seeing the role the duality rotations play emerges from
\newline $~~~~~~~~~~~$
understanding how superspace Lagrangians describe dynamical systems.
\newline $~~~~~~~~~~~$
Every off-shell manifestly 4D, $\cal N$ = 1 supersymmetrical component
level 
\newline $~~~~~~~~~~~$
Lagrangian ${\cal L}{}_{comp}$ can be written in the forms
\be
{\cal L}{}_{comp} ~=~ [\, \rD^a ~,~ \rD^b \, ] [\, \rD_a ~,~ \rD_b \, ] 
{\cal L}{}_{SF}
\ee
$~~~~~~~~~~~$
where ${\cal L}{}_{SF}$ is a superfield Lagrangian expressed in terms
of unconstrained 
\newline $~~~~~~~~~~~~$
prepotentials.  The operator terms $[\, \rD^a ~,~ \rD^b \, ]$ acting on a
monomial involv-
\newline $~~~~~~~~~~~~$
ing a single prepotential superfield will generate a holoraumy containing
an
\newline $~~~~~~~~~~~~$
electromagnetic duality rotation.  In particular, the terms in the purely quad-
\newline $~~~~~~~~~~~~$
ratic portion of any supersymmetical Lagrangian is determined by the
holo-
\newline $~~~~~~~~~~~~$
raumy of the fields in the supermultiplet.  Thus, propagators explicitly
depend
\newline $~~~~~~~~~~~~$
upon the electromagnetic duality rotations.

The equation in (\ref{axROT}) is a pristine example of the power of the 
``SUSY holography'' concept \cite{ENUF} realized between the spaces 
of one dimensional adinkra representations of supersymmetry \cite{Adnk1}
in quantum mechanical systems and spaces of four dimensional representations of 
supersymmetry in field theories.   In the very first introduction of the concept
of 1d, $N$ = 4 holoraumy \cite{1dHolor1}, it was pointed out that the commutator
of supercharges in such systems leads to the appearance of an operator
of the form of a product of a 1d translation times and 1d R-symmetry rotation.
The results expressed in (\ref{axROT}) show that the commutator of supercharges in these 
systems leads to the appearance of an operator of the form of a product of a
translation times an electromagnetic duality rotation. Thus the 1d R-symmetry 
rotation is the ``shadow'' of the 4D electromagnetic duality rotation. 

\vspace{.05in}
 \begin{center}
\parbox{4in}{{\it ``Our knowledge can only be finite, while our ignorance \\ $~~$ 
must necessarily be infinite.'' \\ ${~}$ 
 \\ ${~}$ 
\\ ${~}$ }\,\,-\,\, Karl Popper}
 \parbox{4in}{
 $~~$}  
 \end{center}
 
  \noindent
{\bf Acknowledgements}\\[.1in] \indent
This research of S.\ J.\ G., S.-N.\ Mak, and K.\ S.\ is supported in part by the 
endowment of the Ford Foundation Professorship of Physics at Brown University. 
S.\ J.\ G.\ makes an additional acknowledgment to the National Science 
Foundation grant PHY-1620074.

Additional acknowledgment is given by D.\ L.,  S.-N.\ M.\, and Z.\ W.\ for their 
participation in the second annual Brown University Adinkra Math/Phys Hangout" 
during 2017. 

A final additional acknowledgment is given by D.\ L.\ , B.\ P.\ , A.\ R.\ ,
Z.\ W.\ , X.\ X.\, Y.\ Y.\ , J.\ Z.\, and P.\ V.\ Z.\ for their participation in the 2018 
SSTPRS (Student Summer Theoretical Physics 
Research Session) program at Brown University.


\newpage
\begin{thebibliography}{99}

\bibitem{4dHolor1}
S.\ J.\ Gates, T.\ Grover, M.\ D.\ Miller-Dickson, B.\ A.\ Mondal, A.\ Oskoui, S.\ Regmi, 
E.\ Ross, and R.\ Shetty, ``A Lorentz covariant holoraumy-induced `Gadget' from 
minimal off-shell 4D, N = 1 supermultiplets,'' JHEP {\bf {1511}} (2015) 113,  
e-Print: arXiv:1508.07546 [hep-th], DOI: 10.1007/JHEP11(2015)113.
 
\bibitem{4dHolor2} 
W.\ Caldwell, A.\ N.\ Diaz, I.\ Friend, S.\ J.\ Gates, Jr., S.\ Harmalkar, T.\ 
Lambert-Brown, D.\ Lay, K.\ Martirosova, V.\ A.\ Meszaros, M.\ Omokanwaye, 
S.\ Rudman, D.\ Shin, and A.\ Vershov, ``On the Four Dimensional Holoraumy 
of the 4D, $\cal N$ = 1 Complex Linear Supermultiplet,''  Feb 17, 2017. 27 pp.
Univ. of MD preprint \# PP-017-020, Brown Univ. preprint \# HET-1711,
e-Print: arXiv:1702.05453 [hep-th], DOI: 10.1142/S0217751X18500720.

\bibitem{4dHolor3} 
S.\ J.\ Gates, Jr., S.\ N.\ Hazel Mak.
``Examples of 4D, N = 2 Holoraumy,''
Brown University preprint HET-1770, 
e-Print: arXiv:1808.07946 [hep-th].

\bibitem{1dHolor1} 
S.\ J.\ Gates, Jr., T.\ H\" ubsch, and K.\ Stiffler, 
``Adinkras and SUSY Holography: Some Explicit Examples,'' 
Int.\ J.\ Mod.\ Phys.\ {\bf {A29}} (2014) 07, 1450041, e-Print: arXiv:1208.5999 [hep-th], 
DOI: 10.1142/S0217751X14500419.

\bibitem{1dHolor2} 
S.\ J.\ Gates, Jr., T.\ H\" ubsch, and K.\ Stiffler, 
``On Clifford-algebraic Dimensional Extension 
and  \newline \noindent SUSY Holography,'' 
Int.\ J.\ Mod.\ Phys.\ {\bf {A30}} (2015) 09, 1550042, e-Print: 
arXiv:1409.4445 [hep-th], DOI: 10.1142/S0217751X15500426.

\bibitem{permutadnk} 
I.\ Chappell, II, S.\ J.\ Gates, Jr, and T.\ H\" ubsch, ``Adinkra (In)Equivalence 
From Coxeter Group Representations: A Case Study,''   Int.\ J.\ Mod.\ Phys\. 
{\bf {A29}} (2014) 06, 1450029 e-Print: arXiv:1210.0478 [hep-th], 
DOI: 10.1142/S0217751X14500298.

\bibitem{Frdmn} 
  D.~Z.~Freedman,
  ``Gauge Theories of Antisymmetric Tensor Fields,''
  CALT-68-624.



\bibitem{onSZ}
W. Siegel,
``Off-shell Central Charges,''
Nucl. Phys. B173 (1980) 51-58,
DOI: 10.1016/0550-3213(80)90442-3

\bibitem{Gates:2009me} 
  S.~J.~Gates, Jr., J.~Gonzales, B.~MacGregor, J.~Parker, R.~Polo-Sherk, V.~G.~J.~Rodgers and L.~Wassink,
  ``4D, N = 1 Supersymmetry Genomics (I),''
  JHEP {\bf 0912}, 008 (2009)
  doi:10.1088/1126-6708/2009/12/008
  [arXiv:0902.3830 [hep-th]].
  
  \bibitem{ENUF}
S.\ J. Gates, Jr., W.\ D.\ Linch, III, J. Phillips , ``When Superspace Is Not Enough,'' 
Univ. of Md Preprint \# UMDEPP-02-054, Caltech Preprint \# CALT-68-2387,
arXiv [hep-th:0211034], unpublished.

\bibitem{Adnk1}
M.\ Faux, and S.\ J.\ Gates, Jr.,
``Adinkras: A Graphical technology for supersymmetric representation theory,''
Phys.\ Rev.\ {\bf {D71}} (2005) 065002, e-Print: hep-th/0408004, 
DOI: 10.1103/PhysRevD.71.065002.




   

  
  

  
\end{thebibliography}
\end{document}